\documentclass[aps,prl,twocolumn,amssymb,superscriptaddress]{revtex4-2}

\usepackage{mathtools} 
\usepackage{graphicx}
\usepackage{epstopdf}
\usepackage{amsmath}
\usepackage{dcolumn}
\usepackage{bm}
\usepackage{bbold}
\usepackage{amssymb}
\usepackage{mathrsfs}
\usepackage{amsfonts}
\usepackage{braket}
\usepackage{tabularx}
\usepackage{setspace}
\usepackage[caption=false]{subfig}

\usepackage{float}
\usepackage{array}
\usepackage[dvipsnames]{xcolor}
\usepackage{tablefootnote}
\newcolumntype{L}[1]{>{\raggedright\let\newline\\\arraybackslash\hspace{0pt}}m{#1}}
\newcolumntype{C}[1]{>{\centering\let\newline\\\arraybackslash\hspace{0pt}}m{#1}}
\newcolumntype{R}[1]{>{\raggedleft\let\newline\\\arraybackslash\hspace{0pt}}m{#1}}

\begin{document}
{\normalsize\rightline{DESY-23-084}} 

\preprint{DESY-23-084}

\title{Searching for GUT-scale QCD Axions and Monopoles with a High Voltage Capacitor}

\author{Michael E. Tobar}
\email{michael.tobar@uwa.edu.au}
\affiliation{Quantum Technologies and Dark Matter Labs, Department of Physics, University of Western Australia, 35 Stirling Highway, Crawley, WA 6009, Australia.}
\author{Anton V. Sokolov}
\affiliation{Department of Mathematical Sciences, University of Liverpool, Liverpool, L69 7ZL, United Kingdom}
\author{Andreas Ringwald}
\affiliation{Deutsches Elektronen-Synchrotron DESY, Notkestr. 85, 22607 Hamburg, Germany}
\author{Maxim Goryachev}
\affiliation{Quantum Technologies and Dark Matter Labs, Department of Physics, University of Western Australia, 35 Stirling Highway, Crawley, WA 6009, Australia.}

\begin{abstract}
The QCD axion has been postulated to exist because it solves the strong CP problem. Furthermore, if it exists axions should be created in the early Universe and could account for all the observed dark matter. In particular, axion masses of order $10^{-10}$ to $10^{-7}$ eV correspond to axions in the vicinity of the GUT-scale. In this mass range many experiments have been proposed to search for the axion through the standard QED coupling parameter $g_{a\gamma\gamma}$. Recently axion electrodynamics has been expanded to include two more coupling parameters, $g_{aEM}$ and $g_{aMM}$, which could arise if heavy magnetic monopoles exist. In this work we show that both $g_{aMM}$ and $g_{aEM}$ may be searched for using a high voltage capacitor. Since the experiment is not sensitive to $g_{a\gamma\gamma}$, it gives a new way to search for effects of heavy monopoles if the GUT-scale axion is shown to exist, or to simultaneously search for both the axion and the monopole at the same time.
\end{abstract}


\maketitle

\section{Introduction}

The axion is a putative pseudo-Goldstone boson of Peccei-Quinn (PQ) symmetry breaking, thought to exist because it solves the strong CP problem \cite{PQ1977,Wilczek1978,Weinberg1978,K79,SVZ80,Zhitnitsky:1980tq,DFS81}. Furthermore, the axion is a prime candidate for cold dark matter because it is predicted to be created in the early Universe and can account for all of the observed cold dark matter \cite{Preskill1983,Sikivie1983,Dine1983,Sikivie1983b}. One way the Standard model (SM) particles couple to axions is through the axion-photon chiral anomaly, characterized by the coupling parameter $g_{a\gamma\gamma}$, which is known to modify electrodynamics. Recently the modifications have been expanded to include two other axion-photon anomaly coupling parameters, $g_{aEM}$ and $g_{aMM}$, which occur if magnetic monopoles exist at high energy, as suggested by the theory of Quantum Electromagnetodynamics (QEMD) \cite{Zwanziger1971,SokolovMonopole22,sokolov2023generic}.  

Axion searches typically target the (Kim-Shifman-Vainshtein-Zakharov) KSVZ \cite{K79,SVZ80} and the (Dine-Fischler-Srednicki-Zhitnitsky) DFSZ \cite{Zhitnitsky:1980tq,DFS81} models, where $g_{a \gamma \gamma}=C_{a \gamma \gamma} \alpha /\left(2 \pi f_a\right)$. Here $f_a$ is the high energy scale below the PQ symmetry is broken, with the axion mass given by $m_a\sim5.7\, (10^{15} \text{GeV}/f_a)$ neV \cite{Cortona:2016aa}, where $\alpha$ is the fine structure constant, and $C_{a \gamma \gamma}\sim0.75$ or $-1.92$ for the DFSZ and the KSVZ models, respectively.  
If the PQ symmetry is broken before inflation, light axions can constitute the whole of dark matter, where mass values of $m_a$ between $(0.1-100)~\text{neV}$ correspond to values of $f_a$ near the Grand Unified Theory scale (GUT-scale).

In this work we investigate the sensitivity of a high voltage capacitor to $g_{aEM}$ and $g_{aMM}$ couplings inferred from the resulting axion modified electrodynamics, and conceive viable ways to search for axion dark matter and effects of high energy monopoles at the GUT-scale. Our new proposals could significantly add to the current proposed experimental programs that search for GUT-scale axions via $g_{a \gamma \gamma}$ \cite{ADMXSLIC,ABRACADABRA,FirstAbra,Oue19,ABRA21,GUTDMradio22,Gramolin:2021wm,ADMXSLIC,TOBAR2020,freqmetrology,Thomson:2021wk,Cat21,UpconvCat23,Lasenby2020,Lasenby2020b,berlin2020axion,ABerlin2021,Anyon22}. Here the primary function of the capacitor is to generate a background electric field to gain sensitivity to $g_{aEM}$ and $g_{aMM}$ through an axion generated oscillating electric and magnetic field respectively, which is different to experiments that search for $g_{a\gamma\gamma}$ through electric sensing with a background magnetic field, which have been proposed previously \cite{BEAST,TOBAR2020,Duan23,engelhardt2023detecting,tobar2021abraham}.

\section{Axion-Monopole Modified Electrodynamics with a Static Background Electric Field}

The axion is a pseudoscalar postulated to account for the dark matter halo of our galaxy. Correspondingy, its velocity dispersion is  determined by the galactic virial velocity, $v_a\sim 10^{-3}$, implying a macroscopic de Broglie wave length, 
$\lambda_{\rm dB} = 2\pi/(m_a v_a)\simeq 10^3\,{\rm km}\, ({\rm neV}/m_a)(10^{-3}/v_a)$. Therefore, axion dark matter  
behaves as an approximately spatially homogeneous and monochromatic classical field, which oscillates with a frequency 
determined by the axion mass, $\omega_a\simeq m_a$, and an amplitude proportional to the square root of the energy density of DM in the galactic halo, $\rho_{\rm DM}\simeq 0.45\,{\rm GeV}/{\rm cm}^3$: 
\begin{equation}
a(t,\vec{r}) \simeq \sqrt{2\rho_{\rm DM}} \cos (m_a t)/m_a.
\end{equation}

The generalised axion electrodynamics equations, expanded to include the axion field and the extra coupling terms, $g_{aMM}$ and $g_{aEM}$, in addition to the conventional $g_{a\gamma\gamma}$ term, are given by \cite{sokolov2023generic,SokolovMonopole22,TobarQEMD22,mcallister2022limits,Tong22} (SI units),
\begin{equation}
\begin{aligned}
\vec{\nabla} \cdot \vec{E}_{1}=g_{a\gamma\gamma}c\vec{B}_{0} \cdot \vec{\nabla}a-g_{aEM} \vec{E}_{0} \cdot \vec{\nabla} a+\epsilon_0^{-1}\rho_{e1},
\end{aligned}
\label{GausMP}
\end{equation}
\begin{equation}
\begin{aligned}
\mu_0^{-1}\vec{\nabla} \times \vec{B}_{1}&
=\epsilon_0\partial_t{\vec{E}}_{1}+\vec{J}_{e1}\\
&+g_{a\gamma\gamma}c\epsilon_0\left(-\vec{\nabla} a\times\vec{E}_{0} -\partial_t{a} \vec{B}_{0}\right)\\
&+g_{aEM}\epsilon_0\left(-\vec{\nabla} a\times c^2\vec{B}_{0}+\partial_t{a} \vec{E}_{0}\right),
\end{aligned}
\label{AmpereMP}
\end{equation}
\begin{equation}
\begin{aligned}
\vec{\nabla} \cdot \vec{B}_{1}=-\frac{g_{aMM}}{c} \vec{E}_{0} \cdot \vec{\nabla} a+g_{aEM} \vec{B}_{0} \cdot \vec{\nabla} a,
\end{aligned}
\label{MGaussMP}
\end{equation}
\begin{equation}
\begin{aligned}
\vec{\nabla} \times \vec{E}_{1}&=-\partial_t{\vec{B}}_{1}\\
&+\frac{g_{aMM}}{c}\left(c^2\vec{\nabla} a\times\vec{B}_{0}-\partial_t{a} \vec{E}_{0}\right) \\
&+g_{aEM}\left(\vec{\nabla} a\times\vec{E}_{0}  +\partial_t{a} \vec{B}_{0}\right),
\end{aligned}
\label{FaradayMP}
\end{equation}
where $\vec{B}_{0}$ and $\vec{E}_{0}$ are the impressed background fields, generated from impressed free charge and current densities given by $\rho_{e_0}$ and $\vec{J}_{e_0}$, respectively; $\vec{B}_{1}$ and  $\vec{E}_{1}$ are the axion-induced fields, while $\rho_{e1}$ and $\vec{J}_{e1}$ are the axion-induced charge and current densities.

We assume only a static background electric field so $\vec{B}_{0}=0$ and $\vec{J}_{e_0}=0$, and ignore any possible axion modifications to the background fields. Then the background electric field is determined by the impressed charge density $\rho_{e_0}$, so
\begin{equation}
\begin{aligned}
&\vec{\nabla} \times \vec{E}_0=0\, ,~~~~~\vec{\nabla} \cdot \vec{E}_0=\epsilon_0^{-1}\rho_{e_0}\, .
\end{aligned}
\label{Bground}
\end{equation}
The axion field is strongly repelled by electric charges due to the infinite potential barrier, as has been discussed in Ref.~\cite{Fischler:1983sc} for the dual case of axions interacting with the magnetic charges through the coupling $g_{a\gamma \gamma}$. In particular, in the close $r_0$ vicinity of a charge, the axion field behaves as $a(r) \sim \exp (-r_0/r)$, where $r$ is the radial coordinate associated to a given charge situated at $r=0$. Thus, one has $a=0$ at the locations of the charged particles~\footnote{Note that the couplings $g_{aMM}$ and $g_{aEM}$ arise in the theories which feature heavy magnetic monopoles; in the theories with both electric and magnetic charges, the charged particles in the classical low energy approximation have to be treated as essentially point-like.} and therefore $\vec{\nabla} a \neq 0$ in the vicinity of $r_0$, irrespective of the details of the low energy behaviour of the axion field. Note that this gradient is directed along the electric field $\vec{E}_{0i}$ generated by a given $i$th particle. In contrast, when $r \gg r_0$, the axion field is determined fully by the background, so that $\vec{\nabla} a\sim0$, due to the small $v_a \sim 10^{-3}$ velocities of the DM axions. Thus, importantly in this case one can determine that $\vec{E}_{0} \cdot \vec{\nabla}a=\vec{\nabla}\cdot(a\vec{E}_{0})$, since the product of $a(\vec{\nabla}\cdot\vec{E}_{0})=0$ for all values of $r$ due to the effect described above. Also, since $\vec{\nabla}a$ is along the direction of the electric field, we may conclude $\vec{\nabla}a\times\vec{E}_0=0$. In this case, the axion Maxwell equations (\ref{GausMP})--(\ref{FaradayMP}) with only a static background electric field, become:
\begin{equation}
\begin{aligned}
\vec{\nabla} \cdot (\vec{E}_{1}+g_{aEM}a\vec{E}_{0})=\epsilon_0^{-1}\rho_{e1}\, ,
\end{aligned}
\label{GausMP1}
\end{equation}
\begin{equation}
\begin{aligned}
\mu_0^{-1}\vec{\nabla} \times \vec{B}_{1}&
=\epsilon_0\partial_t({\vec{E}}_{1}+g_{aEM}a\vec{E}_{0})+\vec{J}_{e1}\, ,
\end{aligned}
\label{AmpereMP1}
\end{equation}
\begin{equation}
\begin{aligned}
\vec{\nabla} \cdot (\vec{B}_{1}+\frac{g_{aMM}a\vec{E}_{0}}{c})=0 ,
\end{aligned}
\label{MGaussMP1}
\end{equation}
\begin{equation}
\begin{aligned}
\vec{\nabla} \times \vec{E}_{1}&=-\partial_t({\vec{B}}_{1}+\frac{g_{aMM}a\vec{E}_{0}}{c})\, .
\end{aligned}
\label{FaradayMP1}
\end{equation}

Consequently the effect of DM axions can be described by means of an effective polarization and magnetization \cite{TobarModAx19}  ($\vec{P}_1$ from $g_{aEM}$, and $\vec{M}_1$ from $g_{aMM}$) in the regions where $\vec{E}_0 \neq 0$,  both proportional to $E_0$, and given by:
\begin{eqnarray}
&&\vec{P}_1 = \, g_{aEM}a \epsilon_0\vec{E}_0\, , \\[4pt]
&&\vec{M}_1 = -g_{aMM}ac\epsilon_0\vec{E}_{0} \, .
\end{eqnarray}
This means one can rewrite the Eqs.~(\ref{GausMP1})--(\ref{FaradayMP1}) as follows:
\begin{eqnarray}
&&\vec{\nabla} \cdot \vec{D}_{1}=\rho_{e1}\, , \\[4pt]
\label{GausMP2}
&&\mu_0^{-1}\vec{\nabla} \times \vec{B}_{1}=\partial_t{\vec{D}}_{1}+\vec{J}_{e1} \, , \\[4pt]
\label{AmpereMP2}
&&\vec{\nabla} \cdot \vec{H}_{1}=0 \, , \\[4pt]
\label{MGaussMP2}
&&\vec{\nabla} \times \vec{E}_{1}= -\mu_0\partial_t{\vec{H}}_{1} \, ,
\label{FaradayMP2}
\end{eqnarray}
where the effective auxiliary fields may be defined in vacuo as,
\begin{eqnarray}
&&\vec{D}_1 = \epsilon_0 \vec{E}_1 + \vec{P}_1\, , \\[4pt]
&&\vec{H}_1 = \mu_0^{-1} \vec{B}_1 - \vec{M}_1 \, , \label{AuxField}
\end{eqnarray}

In this work we implement Poynting theorem to calculate the sensitivity of the proposed haloscope. In electrodynamics there are at least four Poynting vectors that may be realised, which over the years lead to the Abraham-Minkowski Poynting theorem controversy \cite{Kinsler_2009}. In this case since the curl of $\vec{M}_{1}$ is zero, then from (\ref{AuxField}), $\vec{\nabla} \times \vec{H}_{1}=\frac{1}{\mu_0}\vec{\nabla} \times \vec{B}_{1}$. Thus, when the background field is a static electric field, the analogous Abraham $(\vec{E}_{1}\times\vec{H}_{1})$ and Minkowski Poynting vectors $(\frac{1}{\epsilon_0\mu_0}\vec{D}_{1}\times\vec{B}_{1})$ are equal \cite{tobar2021abraham}, so for the case of the static electric background any of the four Poynting theorems will give the same result. 

\subsection{Harmonic Equations in Phasor Form}

To solve for harmonic solutions the implementation of the phasor form of Maxwell's equations is a common technique. Here we develop the phasor form of the modified axion electrodynamics. First, the axion pseudo-scalar $a(t)$ may be written as, $a(t)=\frac{1}{2}\left(\tilde{a} e^{-j \omega_a t}+\tilde{a} ^* e^{j \omega_a t}\right)= \operatorname{Re}\left(\tilde{a} e^{-j \omega_a t}\right)$, and thus, in phasor form and in the frequency domain, $\tilde{A}=\tilde{a}e^{-j \omega_a t}$ and $\tilde{A}^*=\tilde{a}^*e^{j \omega_a t}$. In contrast, the electric and magnetic fields as well us the electric current are represented as vector-phasors. For example, we set $\vec{B}_1(\vec{r},t)=\frac{1}{2}\left(\mathbf{B}_1(\vec{r})  e^{-j \omega_1 t}+\mathbf{B}_1^*(\vec{r}) e^{j \omega_1 t}\right)=\operatorname{Re}\left[\mathbf{B}_1(\vec{r}) e^{-j \omega_1 t}\right]$, so we define the vector phasor (bold) and its complex conjugate by, $\tilde{\mathbf{B}}_1(\vec{r},t)=\mathbf{B}_1(\vec{r}) e^{-j \omega_1 t}$ and $\tilde{\mathbf{B}}_1^*(\vec{r},t)=\mathbf{B}_1^*(\vec{r}) e^{j \omega_1 t}$, respectively. Following these definitions, the axion modified Ampere's law in (\ref{AmpereMP1}), and Faraday's law in (\ref{FaradayMP1}), in phasor form become,
\begin{equation}
\begin{aligned}
\frac{1}{\mu_0}\vec{\nabla} \times\mathbf{B}_1=\mathbf{J}_{e1}-j\omega_1\epsilon_0\mathbf{E}_{1}&-j\omega_ag_{aEM}\epsilon_0 \tilde{a}\vec{E}_{0}\, ,
\\
\frac{1}{\mu_0}\vec{\nabla} \times\mathbf{B}_1^{*} =\mathbf{J}_{e1}^*+j\omega_1\epsilon_0\mathbf{E}_{1}^{*}
&+j\omega_ag_{aEM}\epsilon_0\tilde{a}^* \vec{E}_{0} \, ,
\end{aligned}
\label{PhasorAmpE0}
\end{equation}
\begin{equation}
\begin{aligned}
\vec{\nabla} \times \mathbf{E}_1 &=j\omega_1\mathbf{B}_1+j\frac{\omega_ag_{aMM}}{c}\tilde{a}\vec{E}_{0}\, ,
\\
\vec{\nabla} \times \mathbf{E}_1^* &=-j\omega_1\mathbf{B}_1^*-j\frac{\omega_ag_{aMM}}{c}\tilde{a}^*\vec{E}_{0}\, .
\end{aligned}
\label{PhasorFarE0}
\end{equation}
In the following subsections we implement Poynting theorem to and apply it to haloscopes in the reactive regime, well below any resonant frequencies.

\subsubsection{Complex Poynting Theorem}

To implement Poynting theorem to calculate the sensitivity of a reactive system, we need to calculate the imaginary power flow, in a lossless system. For reactive systems the real term can be ignored \cite{tobar2021abraham}, and conversely for resonant systems it is the real power that dominates and the reactive power is ignored. The complex Poynting vector and its complex conjugate are defined by,
\begin{equation}
\mathbf{S}_1=\frac{1}{2\mu_0} \mathbf{E}_1 \times \mathbf{B}_1^{*}~~\text{and}~~\mathbf{S}_1^{*}=\frac{1}{2\mu_0} \mathbf{E}_1^{*} \times \mathbf{B}_1,
\label{AbPv}
\end{equation}
where $\mathbf{S}_1$ is the complex power density of the harmonic electromagnetic wave or oscillation, with the real part equal to the time averaged power density and the imaginary term equal to the reactive power, which may be inductive (magnetic energy dominates) or capacitive (electrical energy dominates). Unambiguously we may calculate the imaginary part of the Poynting vector by,
\begin{equation}
\begin{aligned}
j\operatorname{Im}\left(\mathbf{S}_1\right)=\frac{1}{2}(\mathbf{S}_{1}-\mathbf{S}_{1}^*).
\end{aligned}
\label{ReIm}
\end{equation}Taking the divergence of Eq. (\ref{ReIm}) we find
\begin{equation}
\begin{aligned}
j\vec{\nabla}\cdot\operatorname{Im}\left(\mathbf{S}_{1}\right)=\frac{1}{2}\vec{\nabla}\cdot(\mathbf{S}_{1}-\mathbf{S}_{1}^*)
\end{aligned}
\label{ReImDivEHE0}
\end{equation}
with,
\begin{equation}
\begin{aligned}
&\vec{\nabla} \cdot\mathbf{S}_{1}=\frac{1}{2} \vec{\nabla} \cdot(\mathbf{E}_{1} \times \frac{1}{\mu_0}\mathbf{B}_1^*) =\\
&\frac{1}{2\mu_0}\mathbf{B}_1^* \cdot\vec{\nabla} \times \mathbf{E}_{1}-\mathbf{E}_{1} \cdot\frac{1}{2\mu_0}\vec{\nabla} \times \mathbf{B}_1 \, .
\end{aligned}
\label{DivSEHE0}
\end{equation}
For the reactive solution, $\omega_a=\omega_1$, and substituting Eqs. (\ref{PhasorAmpE0})  and (\ref{PhasorFarE0}) into Eqs. (\ref{DivSEHE0}) and it complex conjugate leads to,
\begin{equation}
\begin{aligned}
&\vec{\nabla} \cdot\mathbf{S}_1=\frac{j\omega_a\epsilon_0}{2}(c^2\mathbf{B}_1^* \cdot\mathbf{B}_1-\mathbf{E}_1 \cdot\mathbf{E}_{1}^{*})-\frac{1}{2}\mathbf{E}_1 \cdot\mathbf{J}_{e1}^* \\
&-\frac{j\omega_a\epsilon_0g_{aEM}}{2}\mathbf{E}_1 \cdot\tilde{a}^* \vec{E}_{0}+\frac{j\omega_a\epsilon_0cg_{aMM}}{2}\mathbf{B}_1^* \cdot\tilde{a}\vec{E}_{0}\, ,
\end{aligned}
\label{divReExEHE0}
\end{equation}
\begin{equation}
\begin{aligned}
&\vec{\nabla} \cdot\mathbf{S}_{1}^*=\frac{j\omega_a\epsilon_0}{2}(\mathbf{E}_1 \cdot\mathbf{E}_{1}^{*}-c^2\mathbf{B}_1^* \cdot\mathbf{B}_1)-\frac{1}{2}\mathbf{E}_1^* \cdot\mathbf{J}_{e1}\\
&+\frac{j\omega_a\epsilon_0g_{aEM}}{2}\mathbf{E}_1^* \cdot\tilde{a} \vec{E}_{0}-\frac{j\omega_a\epsilon_0cg_{aMM}}{2}\mathbf{B}_1 \cdot\tilde{a}^*\vec{E}_{0}\, .
\end{aligned}
\label{divImExEHE0}
\end{equation}
The phase of the axion is not an observable, and is arbitary, so setting $a_0=\tilde{a}=\tilde{a}^*$ and by substituting (\ref{divReExEHE0}) and (\ref{divImExEHE0}) into (\ref{ReImDivEHE0}), as well as realising any induced dissipative electrical currents $\mathbf{J}_{e1}$are proportional to the induced electric fields, $\mathbf{E}_1$ we find,
\begin{equation}
\begin{aligned}
&\vec{\nabla}\cdot\operatorname{Im}\left(\mathbf{S}_{1}\right)=\frac{\omega_a\epsilon_0}{2}(c^2\mathbf{B}_1^* \cdot\mathbf{B}_1-\mathbf{E}_1 \cdot\mathbf{E}_{1}^{*})-\\
&\frac{\omega_a\epsilon_0a_0}{4}\left(g_{aEM}(\mathbf{E}_1^*+\mathbf{E}_1)+g_{aMM}c(\mathbf{B}_1^*+\mathbf{B}_1)\right) \cdot\vec{E}_{0}\, ,
\end{aligned}
\end{equation}
then applying the divergence theorem, we obtain
\begin{equation}
\begin{aligned}
&\frac{\oint\operatorname{Im}\left(\mathbf{S}_{1}\right)\cdot \hat{n}ds}{\omega_a}=\frac{\epsilon_0}{2}\int\Big((c^2\mathbf{B}_1^* \cdot\mathbf{B}_1-\mathbf{E}_1 \cdot\mathbf{E}_{1}^{*})-\\
&\frac{a_0}{2}\left(g_{aEM}(\mathbf{E}_1^*+\mathbf{E}_1)-g_{aMM}c(\mathbf{B}_1^*+\mathbf{B}_1)\right)\cdot \vec{E}_{0}\Big)~dv.
\end{aligned}
\label{PoyntTh}
\end{equation}
If we assume all external reactive sources and sinks are zero, $(\operatorname{Im}\left(\mathbf{S}_{1}\right)\sim0)$ then the reactive power is only supplied by the axion mixing with the static back ground fields, then the reactive stored energy in the circuit may be written as,
\begin{equation}
\begin{aligned}
&U_1=\frac{\epsilon_0}{2}\int\Big((c^2\mathbf{B}_1^* \cdot\mathbf{B}_1-\mathbf{E}_1 \cdot\mathbf{E}_{1}^{*})\Big)~dv\\
&=\frac{\epsilon_0a_0}{4}\int\Big(g_{aEM}(\mathbf{E}_1^*+\mathbf{E}_1)-g_{aMM}c(\mathbf{B}_1^*+\mathbf{B}_1)\Big)\cdot \vec{E}_{0}~dv,
\end{aligned}
\label{Stored}
\end{equation}
where a negative stored energy is capacitive, and a positive stored energy is inductive. Finally from (\ref{Stored}), by squaring the last term and dividing by the second term, we may show that the stored energy may also be expressed as,
\begin{equation}
\begin{aligned}
U_1= \frac{\epsilon_0a_0^2\left(\int\left(g_{aEM}(\mathbf{E}_1^*+\mathbf{E}_1)-g_{aMM}c(\mathbf{B}_1^*+\mathbf{B}_1)\right)\cdot \vec{E}_{0}dv\right)^2}{8\int\Big((c^2\mathbf{B}_1^* \cdot\mathbf{B}_1-\mathbf{E}_1 \cdot\mathbf{E}_{1}^{*})\Big)dv} \, .
\end{aligned}
\label{Stored2}
\end{equation}

\section{High Voltage Capacitor: Sensitivity to Axion-Monopole Couplings}

\subsection{Axion Induced Electric Field}

A high voltage capacitor excited with a static electric background field has been shown to be proportionally sensitive to scalar field dark matter, $\phi(t)$, in the low-mass limit, through the dimensionful coupling constant $g_{\phi\gamma\gamma}$  \cite{Samsonov2022,mcallister2022limits}. For resonant haloscopes, the sensitivity to the axion pseudoscalar field $a(t)$ through the coupling parameter $g_{aEM}$, has been shown to give the same limit as $g_{\phi\gamma\gamma}$ for scalar field dark matter \cite{TobarQEMD22}. Thus, it was hypothesised that a high voltage capacitor may also be sensitive to $g_{aEM}$ in the low mass limit, and here we show that this is indeed true, with the axion dark matter modification to electrodynamics also appearing as an effective polarization. 

From Maxwell electrodynamics it is straightforward to show that the electric field vector phasor inside a cylindrical parallel plate capacitor may be written as \cite{FeynmanCav},
\begin{equation}
\tilde{\mathbf{E}}_1=\tilde{E}_{01}J_0\left(\frac{\omega_a}{c}r\right)e^{-j\omega_at}\hat{z}\, ,~~\tilde{E}_{01}=\frac{\tilde{q}_1}{\epsilon_0\pi R_c^2}=\frac{\tilde{\sigma}_1}{\epsilon_0}\, ,
\label{CapEField}
\end{equation}
and then to confirm the magnetic field vector-phasor as,
\begin{equation}
\tilde{\mathbf{B}}_1=-j\frac{\tilde{E}_{01}}{c}J_1\left(\frac{\omega_a}{c}r\right)e^{-j\omega_at}\hat{\varphi}\, .
\label{CapBField}
\end{equation}
From the series expansion of (\ref{CapEField}) and (\ref{CapBField}), in the quasi-static limit when the Compton wavelength is large compared to the size of the capacitor (as $\omega_a\rightarrow 0$), the magnetic and electric field phasor amplitudes become,
\begin{equation}
\mathbf{E}_1\approx \tilde{E}_{01}\hat{z},~~~~\mathbf{B}_1\approx-j\tilde{E}_{01}\frac{r\omega_a}{2c^2}\hat{\varphi}.
\label{Quasi}
\end{equation}
assuming the electric field is in phase and the magnetic field is out of phase. Given that $\mathbf{B}_1$ is imaginary and $\mathbf{E}_1$ is real, Eq. (\ref{Stored2}) for this experiment becomes,
\begin{equation}
\begin{aligned}
U_1=\frac{g_{aEM}^2a_0^2\epsilon_0\left(\int\mathbf{E}_1\cdot \vec{E}_{0}~dv\right)^2}{2\int\left((c^2\mathbf{B}_1^* \cdot\mathbf{B}_1-\mathbf{E}_1 \cdot\mathbf{E}_{1}^{*})\right)~dv}\, ,
\end{aligned}
\label{StoredCap2}
\end{equation}
which by substituting (\ref{CapEField}) and (\ref{CapBField}) into (\ref{StoredCap2}) gives .
\begin{equation}
\begin{aligned}
U_1=-g_{aEM}^2a_0^2\epsilon_0E_{0}^2Vc\frac{c}{2\pi R_c}\frac{J_1\left(\frac{\omega_a}{c}r\right)}{J_0\left(\frac{\omega_a}{c}r\right)}\, ,
\end{aligned}
\label{StoredCap3}
\end{equation}
where $v_c=\pi R_c^2d_c$ is the volume of the capacitor.
Then, given that $a_0^2=2\langle a_0\rangle^2$, in the quasi static limit ($\omega_a\rightarrow0$), the first term of the expansion of (\ref{StoredCap3}) in powers of $\omega_a$ is a constant term given by,
\begin{equation}
\begin{aligned}
U_1=-g_{aEM}^2\langle a_0\rangle^2\epsilon_0E_{0}^2\pi R_c^2d_c\, .
\end{aligned}
\label{StoredCap4}
\end{equation}
\begin{figure}[t!]
\includegraphics[width=1.0\columnwidth]{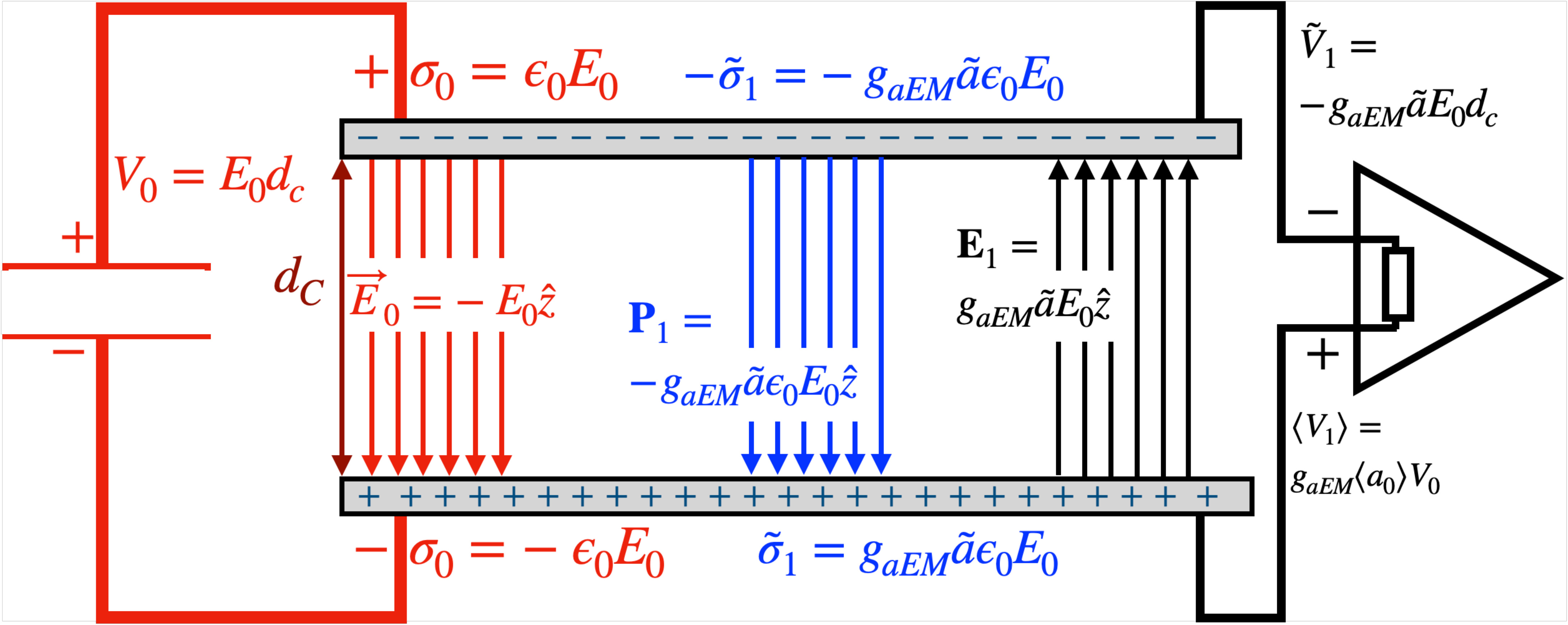}
\caption{Schematic of the proposed experiment, with the capacitor of volume $v_c=\pi R_c^2d_c$, charged by a high-voltage, $V_0$, which produces a static electric field, $\vec{E_0}$, inside the capacitor. Putative axion dark matter interacts with the static field and creates an effective polarization (10) oscillating between the plates, which is discontinuous at the plate boundaries. This produces an alternating voltage measured with the aid of a low noise high-impedance amplifier.}
\label{Capdiag}
\end{figure}
The negative sign just indicates the reactive power delivered to the capacitor is negative. Equating the magnitude of (\ref{StoredCap4}) to the stored energy in the capacitor $|U_1|=\frac{1}{2} \tilde{V_1} \tilde{V_1}^* C_a=\langle V_{1}\rangle^2C_a$, the rms voltage across the capacitor may be calculated as,
\begin{equation}
\begin{aligned}
\langle V_{1}\rangle&=g_{aEM}\langle a_0\rangle V_0,~~\text{where}~~V_0=E_{0}d_c, \\
\text{so}~~~~~\langle E_{01}\rangle&=g_{aEM}\langle a_0\rangle E_{0},
\end{aligned}
\label{rms}
\end{equation}
with a schematic of experiment shown in Fig.\ref{Capdiag}. This is very similar to the proposal to search for the $g_{\phi\gamma\gamma}$ coupling to scalar dark matter \cite{Samsonov2022}.

In the quasi static limit, when $\omega_a^2<\frac{1}{L_1C_1}$, the electrical stored energy is much greater than the magnetic, so (\ref{StoredCap2}) becomes,
\begin{equation}
\begin{aligned}
U_1\approx-\frac{g_{aEM}^2a_0^2\epsilon_0\left(\int\mathbf{E}_1\cdot \vec{E}_{0}~dv\right)^2}{2\int\mathbf{E}_1 \cdot\mathbf{E}_{1}^{*}~dv}.
\end{aligned}
\label{StoredCap5}
\end{equation}
Then substituting in the approximate quasi static electric field from (\ref{Quasi}) into (\ref{StoredCap5}), we may also derive (\ref{StoredCap4}).

The observable for this experiment is the oscillating output voltage, given by
\begin{equation}
\langle V_{1}\rangle=\mathcal{K}_{V_{EM}}g_{aEM}\langle a_0\rangle,~~\text{where}~~\mathcal{K}_{V_{EM}}=V_0,
\label{rms2}
\end{equation}
and $\mathcal{K}_{V_{EM}}$ is the transduction strength from the effective dimensionless axion field ($\theta_{0_{EM}}=g_{aEM}a_0$) to volts \cite{sym14102165}. The spectral noise density, $S_V$, associated with the readout can be measured in units volts squared per Hz, so that the square root spectral density of noise referred to the effective dimensionless axion field may be written as,
\begin{equation}
\sqrt{S_{\theta_{EM}}}=\frac{\sqrt{S_{V}}}{|\mathcal{K}_{V_{EM}}|}.
\label{SSVEM}
\end{equation}
For cold dark matter the signal may be approximated as a narrow band noise source of line width $\Delta f_a$, which is equivalent to an effective Q-factor of $Q_a=\frac{f_a}{\Delta f_a}$. So the signal coherence time is given by, $\tau_a=\frac{Q_a}{f_a}=\frac{1}{\Delta f_a}$. In this case, the signal to noise ratio of the experiment is given by
\begin{equation}
SNR\sim \frac{\mathcal{K}_{V_{EM}}g_{aEM}\left\langle a_{0}\right\rangle}{\sqrt{S_{V}}}\left(t\tau_a\right)^{\frac{1}{4}}=\frac{\left\langle \theta_{0_{EM}}\right\rangle}{\sqrt{S_{\theta_{EM}}}}\left(t\tau_a\right)^{\frac{1}{4}}\, ,
\label{SNRVEM}
\end{equation}
for a measurement time $t>\tau_a$; if $t<\tau_a$, we may substitute $\left(t\tau_a\right)^{\frac{1}{4}}\rightarrow t^{\frac{1}{2}}$, and we use these equations to estimate the sensitivity of this experiment.

\subsection{Axion Induced Magnetic Field Coupled to a Magnetic Circuit} 

\begin{figure}[t!]
\includegraphics[width=1.0\columnwidth]{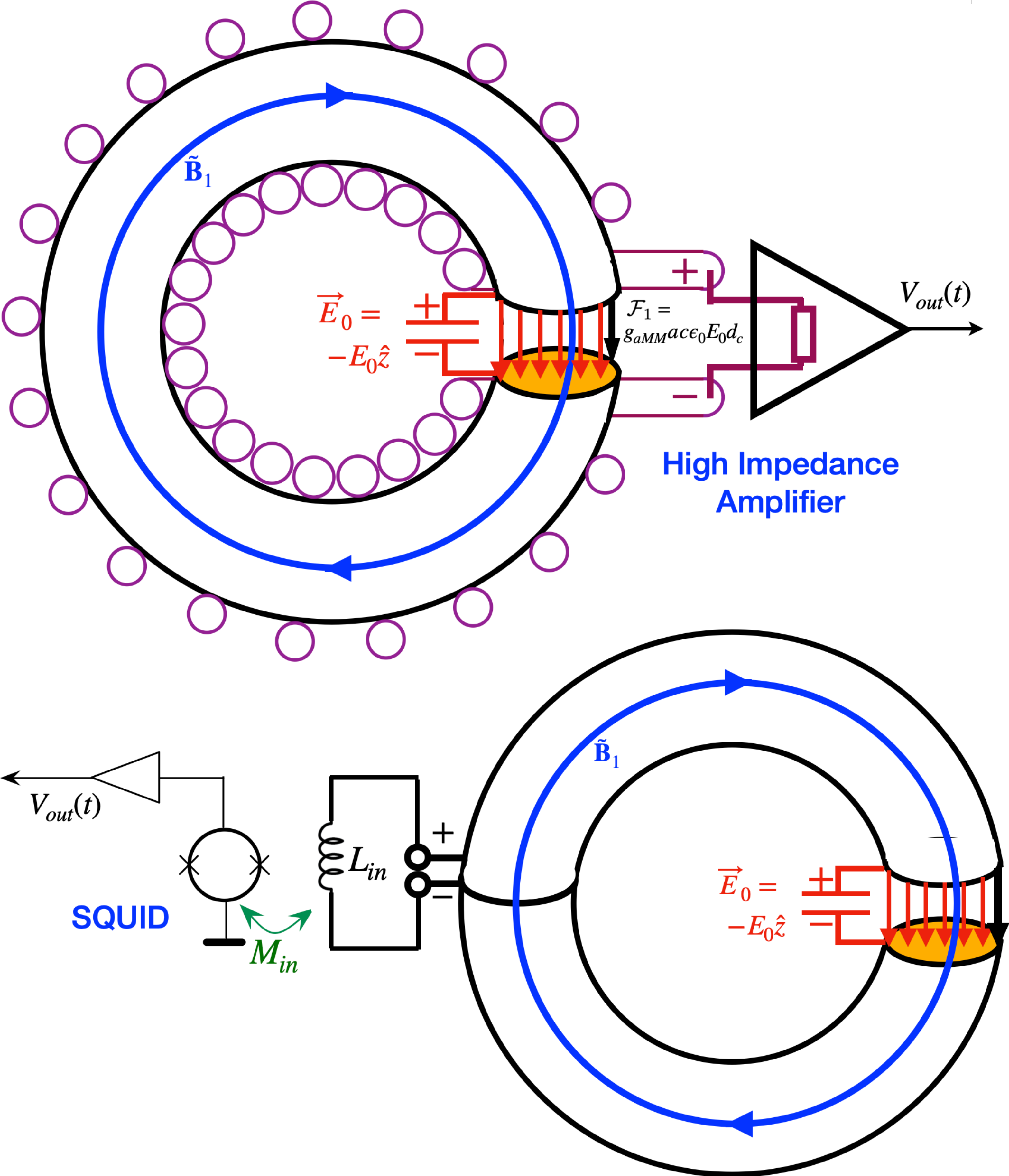}
\caption{Schematic of proposed experiments, with a lossless capacitor of volume $v_c=\pi R_c^2d_c$ charged by a high-voltage and coupled to a toroidal magnetic circuit. The axion interacting with the static electric field produces an oscillating mmf, which generates an oscillating magnetic flux throughout the magnetic circuit. Top, a large winding output read out by a High Impedance Amplifier (HIA). Bottom, a single winding pick up coil readout by a low impedance SQUID amplifier.}
\label{MagC}
\end{figure}

The static electric field produced by the high voltage capacitor will also interact with the axion to create an oscillating magnetic flux density, $\vec{B}_1$, through $g_{aMM}$. To readout $\vec{B}_1$, we may couple the high voltage capacitor to a magnetic circuit as shown in Fig.\ref{MagC}. The magnetic circuit improves the sensitivity in two ways. First, without the magnetic circuit, the axion-induced magnetic field inside and outside the capacitor would be in opposite directions, while the static background electric field will be all in the same direction. This would potentially cause cancellation equivalent to a reduced form factor due to the reduction of the value of $\int\mathbf{B}_1\! \cdot \! \vec{E}_0~dv$. Secondly, without the magnetic circuit the magnetomotive force (mmf) produced by the effective magnetization, $\mathcal{F}_1=\int_{0}^{d_c}\vec{M}_1\cdot \vec{dl}=g_{aMM}ac\epsilon_0E_{0}d_c$, will create a significantly reduced $\vec{B}_1$, because the demagnetization field, $\vec{H}_1$, acts in the opposite direction ($\vec{M}_1=\mu_0^{-1}\vec{B}_1-\vec{H}_1$). Creating a transformer like magnetic circuit readout means the demagnetization field becomes insignificant ($\vec{H}_1\rightarrow0$), so with proper design $\vec{B}_1=\mu_0\vec{M}_1$ within the capacitor.

To construct a low noise readout two approaches may be undertaken as highlighted previously in \cite{TOBAR2020}. The first is to couple to a single loop readout coil, to minimize the readout output impedance and maximize magnetic circuit reluctance, and use a SQUID amplifier in the first stage. The second is to couple to a high inductance coil with multiple windings and readout with a high impedance amplifier in the first stage. The former naturally measures current or flux with a low impedance output, while the latter measures induced voltage with a high impedance output. In both cases we can classify the readouts as impedance mismatch, with the sensitivity determined by the reactive power flow in the circuit.

Assuming the magnetic field is in phase and the induced electric field is out of phase, then we set $\mathbf{B}_1$ as real and $\mathbf{E}_1$ as imaginary, and Eq. (\ref{Stored2}) for this experiment becomes,
\begin{equation}
\begin{aligned}
U_1=\frac{\Big(\frac{g_{aMM}a_0\epsilon_0c}{2}\int\mathbf{B}_1\cdot \vec{E}_{0}~dv\Big)^2}{\int\Big(\frac{1}{2\mu_0}\mathbf{B}_1^* \cdot\mathbf{B}_1
-\frac{\epsilon_0}{2}\mathbf{E}_1 \cdot\mathbf{E}_{1}^{*}\Big)~dv}
\end{aligned}
\label{StoredMC}
\end{equation}
In the quasi static limit, when $\omega_a^2<\frac{1}{L_1C_1}$, the magnetic stored energy is much greater than the electric, so (\ref{StoredMC}) may be approximated as,
\begin{equation}
\begin{aligned}
U_1\approx\frac{g_{aMM}^2a_0^2\epsilon_0}{2}\frac{\Big(\int\mathbf{B}_1\cdot \vec{E}_{0}~dv\Big)^2}{\int\mathbf{B}_1^* \cdot\mathbf{B}_1~dv}\, .
\end{aligned}
\label{StoredMC2}
\end{equation}
The magnetic flux density phasor amplitude is of the form $\mathbf{B}_1\approx-\tilde{B}_{01}e^{-j\omega_at}\hat{\phi}$, and substituting into (\ref{StoredMC2}) we find that,
\begin{equation}
\begin{aligned}
U_1=g_{aMM}^2\langle a_0\rangle^2\epsilon_0E_{0}^2\pi R_c^2d_c\, .
\end{aligned}
\label{StoredMC3}
\end{equation}
Equating the stored energy in (\ref{StoredMC3}) with $\frac{1}{\mu_0}\langle B_{01}\rangle^2 v_c$, the rms amplitude of the magnetic field may be shown to be
\begin{equation}
\begin{aligned}
\langle B_{01}\rangle=g_{aMM}\langle a_0\rangle\frac{E_0}{c}\, .
\end{aligned}
\label{Brms}
\end{equation}

\begin{figure}[t!]
\includegraphics[width=0.3\columnwidth]{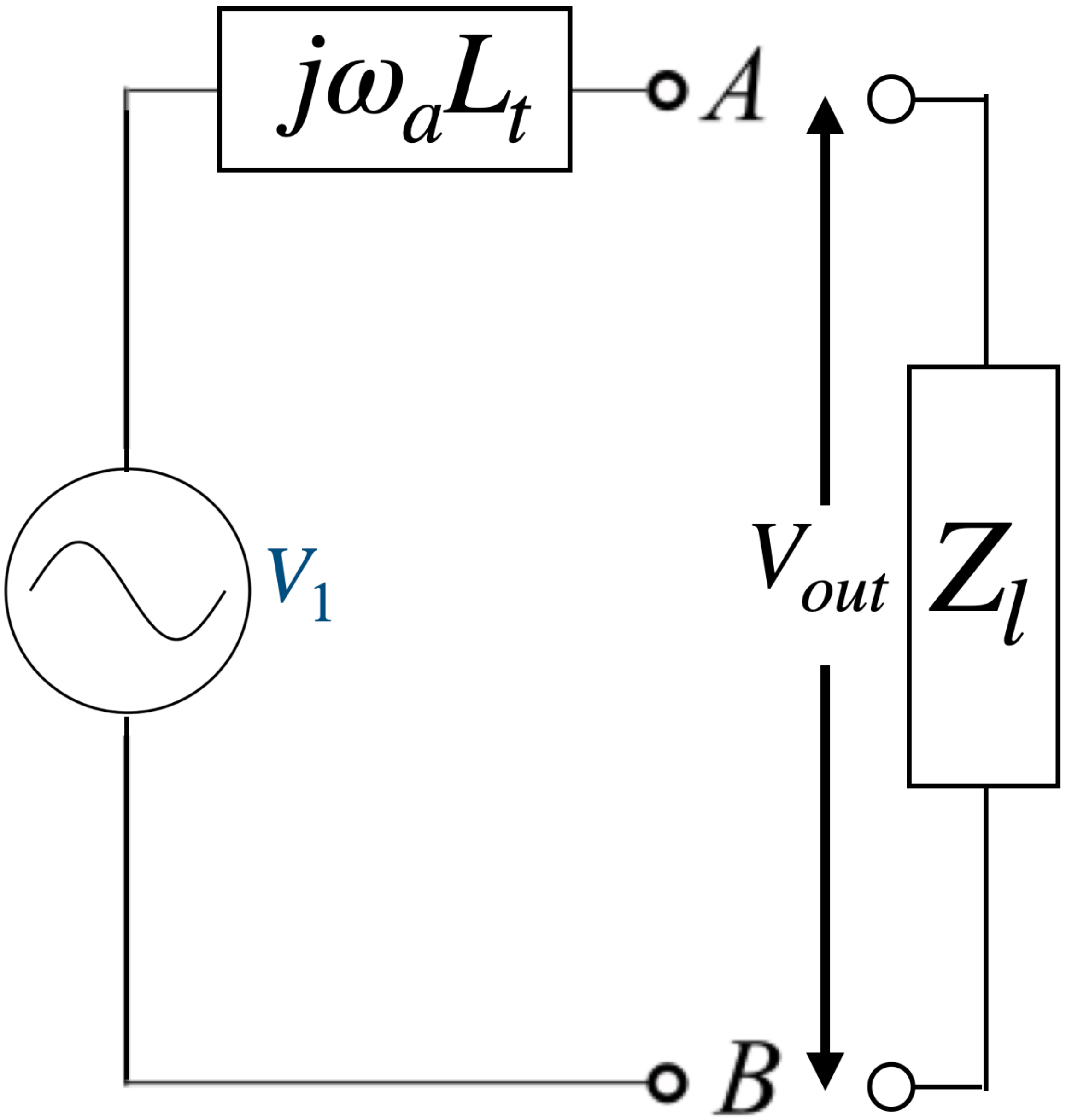}
\caption{Thevenin equivalent circuit at the readout coil of the magnetic circuit shown in Fig.\ref{MagC}. The load impedance $Z_l$ is mainly determined by the input impedance of the SQUID or high impedance amplifier.}
\label{Thev}
\end{figure}

The voltage across the coil is simply given by Faraday's law, $\tilde{V}_{1}=-N_t|\partial_t\mathbf{B}_{1}|\pi R_c^2$, so that the rms output voltage is given by,
\begin{equation}
\begin{aligned}
\langle{V}_{1}\rangle=g_{aMM}\langle a_0\rangle N_t\left(\frac{\omega_a\pi R_c^2}{d_c c}\right)V_0,
\end{aligned}
\label{VoltOut}
\end{equation}
where $N_t$ is the number of turns around the toroidal coil. 
For the toroidal coil with multiple turns coupled to the high impedance amplifier, the inductance is given by, 
\begin{equation}
L_{t}=\frac{\mu_r\mu_0 N_{t}^2\pi R_c^2}{2\pi r_{t}-d_c}, 
\label{Lhigh}
\end{equation}
where $r_t$ the radius to the midpoint of the toroid. For the low impedance output coupled to the SQUID amplifier, the inductance of the single pick up coil is given by, \begin{equation}
L_t\approx\mu_0\mu_r R_s\left[\ln \left(\frac{8 R_c}{r_w}\right)-\frac{7}{4}\right], 
\label{Llow}
\end{equation}
where we assume the coil has a radius $R_c$, and $r_w$ is the radius of the coil wire. The Thevenin equivalent circuit is shown in Fig.\ref{Thev}.

\subsubsection{High Impedance Output} 

The high impedance output requires a large inductance as shown in Fig.\ref{MagC}, this is because the rms value of the axion induced voltage, in (\ref{VoltOut}), is proportional to $N_t$, while the inductance, in (\ref{Lhigh}), $L_t$, is proportional to $N_t^2$. The limit on the value of $L_t$ is set, so it is at least an order of magnitude lower than the input impedance of the high impedance amplifier that reads out the voltage. Typically a high impedance amplifier has an input impedance of order $10~\text{M}\Omega$ \cite{Samsonov2022}, so this gives the restriction of the inductance of the toroid depending on the highest frequency of interest. Assuming $\omega_aL_t<10~\text{M}\Omega$, with the rms voltage in (\ref{VoltOut}) as the observable, we may define the transduction strength as,
\begin{equation}
\begin{aligned}
\mathcal{K}_{V_{MM}}=N_t\left(\frac{\omega_a\pi R_c^2}{d_c c}\right)V_0,
\end{aligned}
\label{KVoltMM}
\end{equation}
so in a similar way to (\ref{SNRVEM}), the signal to noise ratio is given by,
\begin{equation}
SNR\sim \frac{\mathcal{K}_{V_{MM}}g_{aMM}\left\langle a_{0}\right\rangle}{\sqrt{S_{V}}}\left(t\tau_a\right)^{\frac{1}{4}}=\frac{\left\langle \theta_{0_{MM}}\right\rangle}{\sqrt{S_{\theta_{MM}}}}\left(t\tau_a\right)^{\frac{1}{4}},
\label{SNRVMM}
\end{equation}
where $\sqrt{S_{\theta_{MM}}}=\frac{\sqrt{S_{V}}}{|\mathcal{K}_{V_{MM}}|}$.

\subsubsection{Low Impedance Output} 

For the low impedance output we configure the readout with a SQUID amplifier as shown in Fig.\ref{MagC}. In this case the observable can be thought of as magnetic flux created by the axion and picked up by the pick up coil, which senses the induced current. In this case the impedance of the pickup coil should be minimized, so is best realized with a single loop. The rms magnetic flux induced by the axion in the magnetic circuit is given by,
\begin{equation}
\langle\Phi_a\rangle=g_{aMM}\langle a_0\rangle\frac{E_0\pi R_c^2}{c}.
\label{MagFlux}
\end{equation}
The pickup coil of inductance given by (\ref{Llow}), links to the SQUID through a mutual inductance, $M_{in}$ via a SQUID input coil of inductance $L_{in}$, so the SQUID amplifier senses the following magnetic flux,
\begin{equation}
\langle\Phi_{SQ}\rangle=\frac{M_{in}}{L_t+L_{in}}\langle\Phi_{a}\rangle,
\label{MagFluxSQ}
\end{equation}
where, $\Phi_{SQ}$ is our observable so the transduction may be defined as,
\begin{equation}
\begin{aligned}
\mathcal{K}_{\Phi_{MM}}=\frac{M_{in}}{L_t+L_{in}}\frac{\pi R_c^2}{cd_c}V_0,
\end{aligned}
\label{KPhiMM}
\end{equation}
and in a similar way to (\ref{SNRVEM}), the signal to noise ratio is given by,
\begin{equation}
SNR\sim \frac{\mathcal{K}_{\Phi_{MM}}g_{aMM}\left\langle a_{0}\right\rangle}{\sqrt{S_{\Phi_{SQ}}}}\left(t\tau_a\right)^{\frac{1}{4}}=\frac{\left\langle \theta_{0_{MM}}\right\rangle}{\sqrt{S_{\theta_{MM}}}}\left(t\tau_a\right)^{\frac{1}{4}}
\label{SNRIMM}
\end{equation}
where $\sqrt{S_{\theta_{MM}}}=\frac{\sqrt{S_{\Phi_{SQ}}}}{|\mathcal{K}_{\Phi_{MM}}|}$.

\section{Projected Sensitivities}

For the GUT-scale axion the frequencies of interest are considered to be between 24 kHz to 24 MHz, equivalent to an axion mass range between $10^{-10}$ to $10^{-7}$ eV. For the purpose of these calculations we restrict ourselves to the frequency range between 2.4 kHz to 2.4 MHz (between $10^{-11}$ to $10^{-8}$ eV), suitable for the components of the experiment, and still overlapping much of the GUT-scale mass range. For example, many lumped element components do not work well in the MHz range and the low-loss permeable material is only specified up to these frequencies. The experiment will have sensitivity above 2.4 MHz, but it is harder to predict without building, characterizing and calibrating properly at higher frequencies, thus conservatively we just show limits up to 2.4 MHz. 

The experiment may be configured to probe simultaneously $g_{aMM}$ and $g_{aEM}$ by implementing together one of the magnetic circuit readouts to measure the axion induced oscillating magnetic field in (\ref{Brms}) and the high impedance output to measure the axion induced electric field in (\ref{rms}) respectively. Note a limit on $g_{aEM}$ would also give a limit on the scalar field dark matter parameter $g_{\phi\gamma\gamma}$ at the same time \cite{Samsonov2022}.

One can envisage undertaking this experiment with cylindrical oxygen free copper capacitor plates in vacuum. Oxygen free copper has been shown to be able to withstand impulses of electric fields of up to 200 MV/m in vacuum \cite{KEK}. Combining this with the availability of commercially available 600 kV power supplies \cite{Samsonov2022} means that a capacitor plate of 10 cm radius and 0.5 cm separation, would have an electric field of 120 MV/m between the capacitor plates, with a capacitance of $56$ pF. During this work we envisage using such a capacitor cooled to 4~K, with a high impedance amplifier readout over a frequency range of 2.4 kHz to 2.4 MHz, so the impedance of the capacitor would remain about an order of magnitude lower than the input impedance of the low noise high impedance amplifier, which is of the order of $10~\text{M}\Omega$  \cite{Samsonov2022}. Such amplifiers have been shown to have a very low noise spectrum at 4 Kelvin, of order
\begin{equation}
\sqrt{S_V}=\sqrt{\frac{7.4185 \times 10^{-14}}{f^{1.12}}+\frac{9.252 \times 10^{-19}}{f^{0.176}}}~{\rm V}/\sqrt{\rm Hz}\,
\label{NDHIA}
\end{equation}
where $f$ is the Fourier frequency offset in Hz, where we search for the axion at $f=\frac{\omega_a}{2\pi}$.

For the magnetic circuit readout we assume a toroid with a low permittivity core of 14: molypermalloy powder can be used to realise such a core, with efficiency over the frequency range of interest. For the purposes of estimating the sensitivity in a table-top experiment, we assume a $10~$cm scale experiment, with the toroid cross section radius equal to the radius of the capacitor, $R_c=10~$cm. Assuming an average toroidal radius of $45~$cm (similar size to ABRACADABRA 10 cm), we can set the number of turns for the high impedance output to about $600$, so the inductance of the toroid would be, $L_t=70~$mH, giving a maximum impedance of $1~\text{M}\Omega$ at 2.4 MHz. This circuit could be readout with a similar high impedance amplifier with the noise spectrum given by (\ref{NDHIA}).

For the low impedance magnetic circuit readout we want to minimize the inductance, which can be achieved by choosing a relatively thick wire in a single turn pick up coil. If we use a $5$~mm radius wire, the inductance of the pick up coil can be calculated to be $L_t=5.8~\mu \text{H}$. To calculate the noise introduced by the SQUID amplifier, we assume we may reproduce the excellent noise properties of the SHAFT experiment \cite{Gramolin:2021wm}, which we convert to flux noise, and fit to give \cite{sym14102165},
\begin{equation}
\begin{aligned}
&\sqrt{S_{{\phi_{SQ}}}}\sim \\
&\Phi_0\times10^{-6}\sqrt{0.688+\frac{1.76\times10^{30}}{f^{8}}+3.48\times10^{-26}f^4} \\
&{\rm Wb}/\sqrt{\rm Hz},
\end{aligned}
\label{NDSQ}
\end{equation}
where $\Phi_0=\frac{h}{2e}=2.0678\times10^{-15}$ Wb is the magnetic flux quantum. Typical SQUID parameters set $M_{in}\sim 8~\text{nH}$ and $L_{in}\sim150~\text{nH}$ \cite{Gramolin:2021wm,TOBAR2020}, which we use in our sensitivity calculations.

There may be a question on how the low noise readouts perform in the presence of such a large, applied DC voltage and electric field. First, the SQUID and high impedance amplifier circuits that readout and search for $g_{aMM}$ couplings may be electrically isolated from the high voltage and fields, as a non-conducting element may be place between the magnetic circuit and capacitor plate, and the SQUID or high impedance amplifier may be far away from the fringing electric fields.  Secondly, the high impedance amplifier that measures the AC electric field effects and searches for $g_{aEM}$ couplings must be AC coupled, so the DC voltage is suppressed at the input, and it would have to be properly designed and characterised for the frequency range of interest. This situation can occur in ion trapping experiments, for which the high impedance amplifier was designed for, and should be able to be solved so an extremely sensitive search may be undertaken.

\begin{figure}[t!]
\includegraphics[width=1.0\columnwidth]{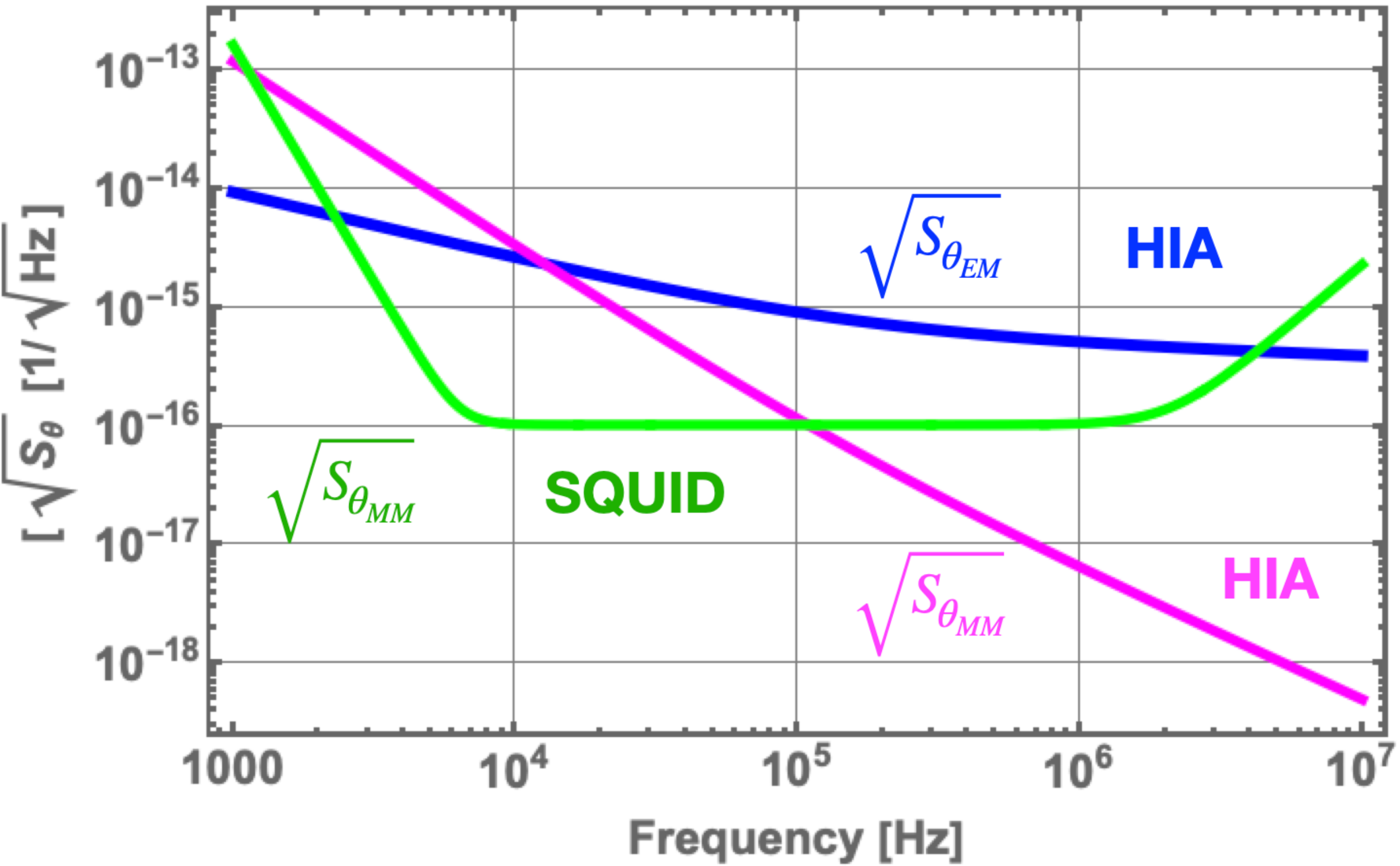}
\caption{Estimated spectral sensitivity for the three proposed detectors, from equations (\ref{SSVEM}), (\ref{SNRVMM}) and (\ref{SNRIMM}), using the assumed parameters given in the text.}
\label{SpecSens}
\end{figure}

\begin{figure}[t!]
\includegraphics[width=1.0\columnwidth]{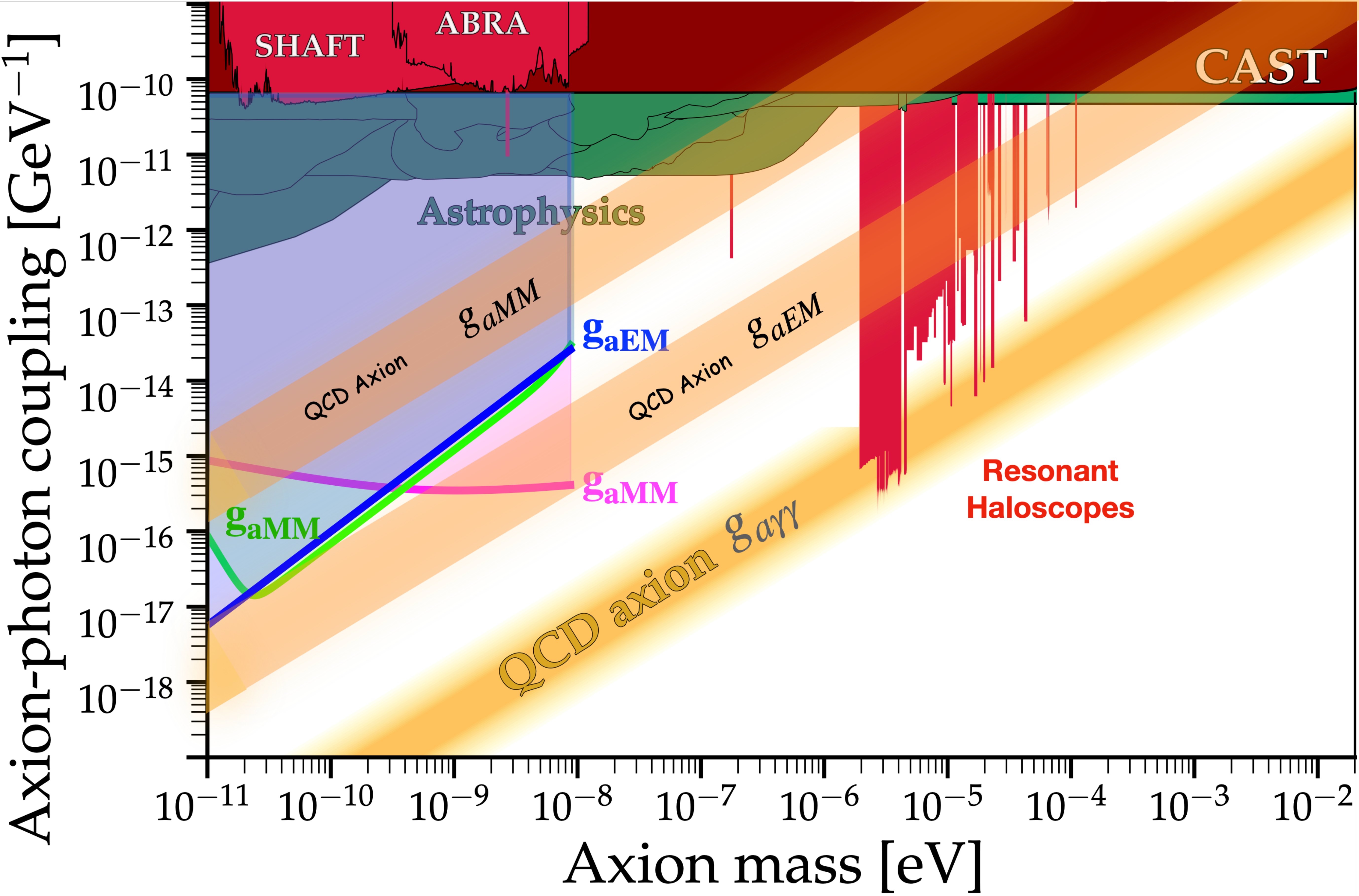}
\caption{Estimated order of magnitude sensitivity to axion-photon coupling parameters, $g_{aEM}$ and $g_{aMM}$, for the three proposed detectors, derived from equations (\ref{SNRVEM}), (\ref{SNRVMM}) and (\ref{SNRIMM}), with SNR set to unity, using the assumed parameters given in the text and with 18 days of continuous data taking. Note the resonant haloscopes \cite{hagmann1990,Hagmann1990b,Bradley2003,ADMXaxions2010,ADMX2011,Braine2020,Bartram2021,McAllister2017,Quiskamp2022,universe7070236,10.1093/ptep/ptab051,Devlin2021,Kwon2021,Backes:2021wd,IWAZAKI2021115298,Chigusa:2021vh,PhysRevD.103.096001,Alvarez-Melcon:2021aa,Alesini22}, SHAFT~\cite{Gramolin:2021wm}, ABRA~\cite{ABRA21} and ADMX SLIC~\cite{ADMXSLIC} are only sensitive to $g_{a\gamma\gamma}$ and not to $g_{aMM}$ and $g_{aEM}$. The constraints from astrophysics and CAST hold for both the $g_{a\gamma \gamma}$ and the $g_{aMM}$ couplings~\cite{SokolovMonopole22}. Moreover, CAST \cite{Anastassopoulos2017} has been shown to have sensitivity to $g_{\phi\gamma\gamma}$ and hence $g_{aEM}$ \cite{Samsonov2022}. The figure has been adapted from limits listed in the reference \cite{AxionLimits}.}
\label{Sens}
\end{figure}

First we present the sensitivities in terms of the effective dimensionless axion spectral noise, which is independent of the axion signal and only depends on the transduction sensitivity and noise in the detector \cite{sym14102165}, and is plotted in Fig.\ref{SpecSens}. Following this, we assume that putative axions make up all of the galactic halo dark matter density, and present as a narrow band noise source due to virialization of dark matter within the halo. For this type of signal the signal to noise ratios are given by equations (\ref{SNRVEM}), (\ref{SNRVMM}) and (\ref{SNRIMM}), where the rms value of the effective dimensionless axion field is related to the dark matter density, $\rho_{DM}$, by,
\begin{equation}
\left\langle \theta_{0_i}\right\rangle=g_{ai}\frac{\sqrt{\rho_{D M} c^{3}}}{\omega_{a}} \, ,
\end{equation}
where $i=MM$ or $EM$. The order of magnitude exclusion limits are set by assuming SNR=1 and assuming the experiment runs for 18 days continuously, and are plotted in Fig.~\ref{Sens}. From the plots we may conclude that these experiments can search for the GUT-scale QCD axion if putative heavy monopoles exist. \\

\noindent\textbf{Acknowledgments}

This work was funded by the Australian Research Council Centre of Excellence for Engineered Quantum Systems, CE170100009 and the Australian Research Council Centre of Excellence for Dark Matter Particle Physics, CE200100008. AVS is funded by the UK Research and Innovation grant MR/V024566/1. AR acknowledges support by the Deutsche Forschungsgemeinschaft (DFG, German Research Foundation) under Germany’s Excellence Strategy - EXC 2121 Quantum Universe - 390833306 and under - 491245950.

\providecommand{\noopsort}[1]{}\providecommand{\singleletter}[1]{#1}%

\end{document}